\documentclass[12pt]{amsart}

\usepackage{amscd}
\usepackage{amsmath}
\usepackage{graphicx}
\usepackage{amsfonts}
\usepackage{amssymb}
\begin{document}
\title[Uncertainty relations]
{Uncertainty relations for any multi observables}
\author{Jinchuan Hou, Kan He}
\address {College of
Mathematics, Institute of Mathematics, Taiyuan University of
Technology, Taiyuan 030024, P. R.  China}
 \email[J.-C. Hou]{jinchuanhou@aliyun.com}\email[K.
He]{hekan@tyut.edu.cn}

\thanks{{\it PACS 2010.}  03.65.Ta, 03.65.Db, 02.30.Tb}

\begin{abstract}

 Uncertainty  relations describe  the lower bound of product of standard
deviations of observables.  By revealing a  connection between
standard deviations of quantum observables and numerical radius of
operators, we  establish  a universal uncertainty relation for any
$k$ observables, of which the formulation depends on the even or odd
quality of $k$. This universal uncertainty relation is tight at
least for the cases $k=2n$ and $k=3$. For two observables, the
uncertainty relation is exactly a simpler reformulation of
Schr\"{o}dinger's uncertainty principle.

\end{abstract}
\maketitle

\section{Introduction}

In the past  ninety years, the theory of quantum mechanics was
applied in lots of other sciences, including Information Science,
Chemistry and Biology (Ref. \cite{BD,Eng,LW}).  The uncertainty
principle, discovered first by Heisenberg in 1927 (Ref. \cite{Hei}),
is often considered as one of the most important topics of quantum
theory (Ref. \cite{Hei2,Von}) and can be linked to quantum
entanglement and other important topics (Ref. \cite{BC,Pre}).
Heisenberg's uncertainty principle says that
$$ \Delta_q \Delta_p\geq \frac{\hbar}{2},\eqno(1.1)$$
where $\Delta_q$ and $ \Delta_p$ denote standard deviations of the
position $\hat{q}$ and momentum $\hat{p}$ respectively,
$\hbar=\frac{1}{2}|\langle \hat{q}\hat{p}-\hat{p}\hat{q}\rangle|$ is
the reduced Planck constant.
 Recalled that a quantum system
can be simulated in a complex Hilbert space $H$ with the inner
product $\langle\cdot|\cdot\rangle$ and a pure state is described by
a unit vector
 $|x\rangle$. Quantum  observables for a state $|x\rangle$ are  self-adjoint operators  on $H$ with domain containing $|x\rangle$ (Ref. \cite{Von}).
The value of observable $A$ for the pure state $|x\rangle$ is
$\langle A\rangle=\langle x|A|x\rangle$. In 1929, Robertson
generalized Heisenberg's uncertainty principle, which says that, for
observables $A, B$ and  pure state $|x\rangle$, $$ \Delta_A
\Delta_B\geq \frac{1}{2} |\langle [A,B]\rangle|.\eqno(1.2)$$ Where
$[A,B]=AB-BA$ is the Lie product of $A$ and $B$,
$$\Delta_A=\sqrt{\langle A^2\rangle-\langle A\rangle^2} \ {\rm and}\
\Delta_B=\sqrt{\langle B^2\rangle-\langle B \rangle^2}$$ are the
standard deviations of $A$ and $B$, respectively \cite{Rob}.
Schr\"{o}dinger gave a  uncertainty principle, which is sharper than
Robertson's and asserts that $$\Delta_A  \Delta_B \geq
\sqrt{\frac{1}{4} |\langle [A,B]\rangle|^2+ |\frac{1}{2}\langle
\{A,B\}\rangle-\langle A\rangle \langle B\rangle |^2},\eqno(1.3)$$
where $\{A,B\}=AB+BA$ is the Jordan product of $A$ and $B$
\cite{Sch}. Schr\"{o}dinger's uncertainty principle holds for mixed
state, too. Recall that a mixed state $\rho$ is a positive operator
on $H$ with trace 1. Then the value of observable $A$ for the state
$\rho$ is $\langle A\rangle={\rm Tr} (A\rho)$ and the standard
deviation of $A$ is $\Delta_A=\sqrt{\langle A^2\rangle-\langle
A\rangle^2}=\sqrt{{\rm Tr}(A^2\rho)-{\rm Tr}(A\rho)^2}$. Here we
assume that both ${\rm Tr}(A\rho)$ and ${\rm Tr}(A^2\rho)$ are
finite.

What   happens for multi observables?

There is a simple way to get certain uncertainty relation from the
uncertainty principles in (1.2) or (1.3). For example, let $A,B,C$
be three observables, then by applying (2) one gets
$$\Delta_A^2\Delta_B^2\Delta_C^2\geq \frac{1}{8}|\langle
[A,B]\rangle\langle [B,C]\rangle\langle [A,C]\rangle|.\eqno(1.4)$$
But (1.4) is not sharp enough.

Let $A=\hat{q}, B=\hat{p}$ and $C=\hat{r}=-\hat{p}-\hat{q}$; then
$[p,q]=[q,r]=[r,p]=\frac{\hbar}{i}$, and thus  (1.4), together with
(1.1), gives
$$\Delta_q^2\Delta_p^2\Delta_r^2\geq(\frac{\hbar}{2}) ^3.$$
However, in \cite{KW}, a tight uncertainty relation is given that $$
\Delta_q^2\Delta_p^2\Delta_r^2\geq (\tau \frac{\hbar}{2}) ^3
\eqno(1.5)$$ with $\tau=\frac{2}{\sqrt{3}}>1$.

This also happens for  Pauli matrices $X,Y,Z$. As $[X,Y]=2iZ$, one
has $\frac{1}{2}|\langle [X,Y]\rangle|= |\langle Z\rangle|$.
Similarly
 $\frac{1}{2}|\langle [X,Z]\rangle|= |\langle Y\rangle|$ and
$\frac{1}{2}|\langle [Y,Z]\rangle|= |\langle X\rangle|$. Thus by
(1.4)
$$ \Delta_X^2\Delta_Y^2\Delta_Z^2\geq  |\langle
X\rangle\langle Y\rangle\langle Z\rangle|.$$ But it was announced by
S.-M. Fei  that
$$
\Delta_X^2\Delta_Y^2\Delta_Z^2\geq \frac{8}{3\sqrt {3}}{|\langle
X\rangle\langle Y\rangle\langle Z\rangle|}. \eqno(1.6)$$ This
inequality is also tight and achieves ``=" at
$\rho=\frac{1}{2}(I+\frac{1}{\sqrt{3}}X+\frac{1}{\sqrt{3}}Y+\frac{1}{\sqrt{3}}Z)$.

Therefore, to obtain uncertainty relations for multi observables
that are sharp enough, one needs new approaches. Let $A_1,A_2,
\ldots, A_k$ be any $k$ observerbles of a quantum system. The
purpose of this paper is to establish a lower bound of
$\Delta_{A_1}\Delta_{A_2}\cdots\Delta_{A_k}$ in terms of $\langle
A_iA_j\rangle, \langle A_i\rangle\langle A_j\rangle$ and $\langle
A_j^2\rangle$.  For the case when $k=2$, the uncertainty relation is
equivalent to Schr\"{o}dinger's uncertainty principle that is tight
and has a simpler representation. For the case when $k=3$, we show
that the untainty relation is tight by taking Pauli matrices as
observables. All proofs of the main result and  the lemmas will be
presented in the appendix section.

\section {uncertainty relations for multi observables}

 Our main idea is based on the following observation, which establishes a formula to connect the standard deviation of  a
  quantum observable $A$ of a state $|x\rangle$ to the norm as well as the numerical radius of  $[A,|x\rangle\langle x|]$,
  the Lie product of $A$ and the rank one projection $|x\rangle\langle x|$.

 Let $T$ be a bounded linear operator acting on a complex Hilbert
space $H$. The numerical range of $T$ is the set $W(T)=\{\langle
x|T|x\rangle \,:\, |x\rangle\in H, \||x\rangle\|=1 \},$
 and the numerical radius of $T$ is $w(T)=\sup\{|\lambda| \,:\, \lambda \in
 W(T)\}.$ The topic of numerical range and numerical radius plays an important role in mathematics and is applied into many
 areas (Ref. \cite{GR, halmos-book,HHZ}). Denote by $\|T\|$  the operator norm of $T$.

 {\bf Lemma 2.1}. {\it Let  $|x\rangle$ be a pure state and $A$
an observable for it. \if Let $H$ be a complex Hilbert space, $w$
the numerical radius, and $|x\rangle$ the pure state  belongs to the
domain of observable $A$.\fi Then
$$ \Delta_A=\|[A,|x\rangle\langle x|]\|=w([A, |x\rangle\langle
x|]).$$}

By Lemma 2.1, for any observables $A_1,A_2,\ldots, A_k$ for a pure
state $|x\rangle$,
$$ \Pi_{j=1}^k\Delta_{A_j}=\Pi_{j=1}^k\|[A_j,|x\rangle\langle
x|]\|\geq \|\Pi_{j=1}^k [A_j,|x\rangle\langle x|]\|\geq
w(\Pi_{j=1}^k [A_j,|x\rangle\langle x|]). \eqno(2.1)
$$
Note that, the value of $ \Pi_{j=1}^k\Delta_{A_j}$ does not  depend
on the order arrange of observables but $w(\Pi_{j=1}^k
[A_j,|x\rangle\langle x|])$ does. Therefore, the  inequality (2.1)
can be sharped to
$$ \Pi_{j=1}^k\Delta_{A_j} \geq \max_\pi
w(\Pi_{j=1}^k [A_{\pi(j)},|x\rangle\langle x|]), \eqno(2.2)
$$
where the maximum is over all permutations $\pi$ of $(1,2,\ldots,
k)$. Thus the question of establishing an uncertainty relation for
$k$ observales is reduced to the question of calculating the
numerical radius of the operator
$$D_k^{(\pi)}=\Pi_{j=1}^k [A_{\pi(j)},|x\rangle\langle x|], \eqno(2.3)$$
which is an operator of rank $\leq 2$.

The exact value of $w(D_k^{(\pi)})$ is computable and we can
establish an  uncertainty relation for any multi observables by
(2.2). For simplicity, and with no loss of generality, we state our
  results only for $\pi={\rm id}$.

  One may ask why do not work on the stronger inequality $$ \Pi_{j=1}^k\Delta_{A_j}\geq \|\Pi_{j=1}^k [A_j,|x\rangle\langle x|]\|?$$
In fact, as we show in the Appendix section, this stronger
inequality leads to weaker uncertainty relations. So the numerical
radius is the better choice.

The following is our main result, here we agree on
$\Pi_{j\in\Lambda} a_j=1$ if $\Lambda=\emptyset$. It is surprising
that our uncertainty relation for any $k$ observables has  different
formulation depending on the even or odd quality of the integer $k$.

{\bf Theorem 2.2}. {\it Let $A_1,A_2,\ldots, A_k$ with $k\geq 2$ be
observables.

{\rm (1)} If $k=2n$, then
$$\begin{array}{rl} &
\Pi_{j=1}^{2n}\Delta_{A_j}
\\ \geq &\frac{1}{2}(\Pi_{j=1}^{n-1}|\langle
A_{2j}A_{2j+1}\rangle-\langle A_{2j } \rangle\langle
 A_{2j+1}\rangle|)(|\langle A_1A_{2n}\rangle-\langle A_1\rangle\langle
A_{2n}\rangle|+\Delta_{A_1}\Delta_{A_{2n}}).
\end{array}\eqno(2.4)$$

{\rm (2)} If $k=2n+1$, then, identifying $2n+2$  with $[(2n+2)$ {\rm
mod} $(2n+1)]=1$,
$$\begin{array}{rl} &
\Pi_{j=1}^{2n+1}\Delta_{A_j}
\\ \geq &\frac{1}{2}[ 2\Pi_{j=1}^{2n+1}|\langle A_{j}A_{j+1}\rangle-\langle
A_{j}\rangle\langle A_{j+1}\rangle|\\
& + \Delta_{A_1}^2 \Pi_{j=1}^{n}|\langle
A_{2j}A_{2j+1}\rangle-\langle A_{2j}\rangle\langle
A_{2j+1}\rangle|^2
\\ & +\Delta_{A_{2n+1}}^2\Pi_{j=1}^{n}|\langle
A_{2j-1}A_{2j}\rangle-\langle A_{2j-1}\rangle\langle
A_{2j}\rangle|^2]^{\frac{1}{2}}.
\end{array} \eqno(2.5)$$
}

Obviously, ``=" holds if and only if
$$\Pi_{j=1}^k\|[A_j,|x\rangle\langle
x|]\|= \|\Pi_{j=1}^k [A_j,|x\rangle\langle x|]\|= w(\Pi_{j=1}^k
[A_j,|x\rangle\langle x|]). \eqno(2.6)
$$
Thus the uncertainty relation is tight if Eq.(2.6) holds for some
observerbles $A_1,A_2,\ldots, A_k$ and some state. This is the case
as will be illustrated in Section 4.

 We remak that Theorem 2.2 holds
for any state $\rho$ with $|{\rm Tr}(A_j\rho)|<\infty$ and ${\rm
Tr}(A_j^2\rho)<\infty$, $j=1,2,\ldots,k$.  To see this, denote by
${\mathcal C}_2(H)$ be the Hilbert-Schimit class in $H$, which is a
Hilbert space with inner product $\langle T,S\rangle={\rm Tr}(T^\dag
S)$. Then, a positive operator $\rho$ is a state if and only if
$\sqrt{\rho}$ is a unit vector in ${\mathcal C}_2(H)$. For a
self-adjoint operator $A$ on $H$, define a linear operator $L_A$ on
${\mathcal C}_2(H)$ by $L_AT=AT$ if ${\rm Tr}(T^\dag A^2T)< \infty$.
It is clear that $L_A$ is self-adjoint as $L_A^\dag=L_{A^\dag}=L_A$.
Note that
$$\langle A\rangle={\rm Tr}(A\rho)=\langle
\sqrt{\rho}|L_A|\sqrt{\rho}\rangle=\langle L_A\rangle$$ and thus
$$\Delta_A=\Delta _{L_A}, \quad \langle L_AL_B\rangle=\langle L_{AB}\rangle=\langle AB\rangle.$$
Then, Theorem 2.2 is true by applying (2.1) to $L_{A_1},
L_{A_2},\ldots, L_{A_k}$ and the pure state $|\sqrt{\rho}\rangle$.

Before to see the uncertainty relations presented by theorem 2.2 is
sharper than those obtained by Heisenberg's uncertainty principle
(1.2) and Schr\"{o}dinger's uncertainty principle (1.3), we
illustrate some application of Theorem 2.2 for the cases $k=2$.

\section{The case of $k=2$: a reformulation of Schr\"{o}dinger's
principle}

Applying Theorem 2.2 (1) to the case when $k=2$, the following
result is immediate.

{\bf Theorem 3.1}. {\it Let    $A$ and $B$ be  observables for a
state. Then $$ \Delta_A\Delta_B\geq |\langle AB \rangle-\langle
A\rangle \langle B \rangle|, \eqno(3.1)$$  which is equivalent to
Schr\"{o}dinger's uncertainty principle.}

The expression of inequality is quite simpler than that of
Schr\"{o}dinger's uncertainty principle. We show that (3.1) is in
fact equivalent to Schr\"{o}dinger's uncertainty principle (1.3).

To check it, write $\langle A\rangle \langle B \rangle=r$ and
$\langle AB \rangle=s+ it$, where $s,t,r\in \Bbb R$. Then $\langle
BA\rangle=s- it$. A simple computation gives
$$|\langle AB \rangle-\langle
A\rangle \langle B \rangle|=\sqrt{(s-r)^2 +t^2},$$
$$\sqrt{\frac{1}{4} |\langle [A,B]\rangle|^2+ |\frac{1}{2}\langle
\{A,B\}\rangle-\langle A\rangle  \langle B\rangle |^2}=\sqrt{(s-r)^2
+t^2}$$ and $$\frac{1}{2} |\langle [A,B]\rangle|=|t|.$$ So,  we get
$$\begin{array}{lll}  \Delta_A\Delta_B & \geq  |\langle A\rangle
\langle B \rangle -\langle AB \rangle| \\ & =   \sqrt{\frac{1}{4}
|\langle [A,B]\rangle|^2+ |\frac{1}{2}\langle \{A,B\}\rangle-\langle
A\rangle  \langle B\rangle |^2} \\ & \geq \frac{1}{2}|\langle
[A,B]\rangle|.\end{array}\eqno(3.2)$$

 Now we are at a position to show that Theorem 2.2 is sharper than
 the uncertainty relations obtained by the approach mentioned in the
 introduction section.

 Let $A_1,A_2,\ldots, A_k$   be observables.

 If $k=2n$ is even, by inequalities
 (3.2) one has
 $$\begin{array}{rl} \Pi_{j=1}^{k}\Delta_{A_j}= &
 (\Pi_{j=1}^{n-1}(\Delta_{A_{2j }}\Delta_{A_{2j+1}}))(\Delta_{A_{1}}\Delta_{A_{2n}})\\
 \geq & (\Pi_{j=1}^{n-1} |\langle A_{2j }A_{2j+1}\rangle-\langle A_{2j }\rangle\langle
 A_{2j+1}\rangle|)\langle A_1A_{2n}\rangle-\langle
A_1\rangle\langle A_{2n}\rangle| \\
 \geq & \frac{1}{2^n}(\Pi_{j=1}^{n-1} |\langle[A_{2j }, A_{2j+1}]\rangle|)|\langle[A_{1}, A_{2n}]\rangle|,
 \end{array} \eqno(3.3)
 $$
which is weaker than the inequality (2.4) since
$\Delta_{A_{1}}\Delta_{A_{2n}}\geq |\langle A_1A_{2n}\rangle-\langle
A_1\rangle\langle A_{2n}\rangle|$.

If $k=2n+1$ is odd, by (3.2) again we have
$$\begin{array}{rl} \Pi_{j=1}^{k}\Delta_{A_j}^2= &
 (\Pi_{j=1}^n(\Delta_{A_{2j-1}}\Delta_{A_{2j}}))( \Pi_{j=1}^n(\Delta_{A_2j }\Delta_{A_{2j+1}}))(\Delta_{A_1 }\Delta_{A_{2n+1}})\\
 \geq & (\Pi_{j=1}^n |(\langle A_{2j-1}A_{2j}\rangle-\langle A_{2j-1}\rangle\langle
 A_{2j}\rangle)(\langle A_{2j }A_{2j+1}\rangle-\langle A_{2j }\rangle\langle
 A_{2j+1}\rangle)|)\cdot \\ & \cdot |\langle A_{1 }A_{2n+1}\rangle-\langle A_{1 }\rangle\langle
 A_{2n+1}\rangle| \\
 \geq & \frac{1}{2^{2(2n+1)}}(\Pi_{j=1}^n ( |\langle[A_{2j-1}, A_{2j}]\rangle\langle [A_{2j},A_{2j+1}]\rangle|))|\langle [A_1,A_{2n+1}]\rangle|,
 \end{array} \eqno(3.4)
 $$
which is clearly weaker than the inequality (2.5) as $a^2+b^2\geq 2
ab$ and $\Delta_{A_{1}}\Delta_{A_{2n+1}}\geq |\langle
A_1A_{2n+1}\rangle-\langle A_1\rangle\langle A_{2n+1}\rangle|$.

\section{Uncertainty relations for three or four observables}

By Theorem 2.2 and a careful check of its proof, one gets a
uncertainty relation for any three observables like the following.

{\bf Theorem 4.1}. {\it Let $A,B,C$ be three observables for a state
$\rho$ in a state space $H$, then
$$ \begin{array}{rl}
\Delta_A^2\Delta_B^2\Delta_C^2\geq & \frac{1}{4}(\Delta_C^2|\langle
AB\rangle-\langle A\rangle\langle B\rangle|^2+\Delta_A^2|\langle
BC\rangle-\langle B\rangle\langle C\rangle|^2 )\\
& +\frac{1}{2}|(\langle AB\rangle-\langle A\rangle\langle
B\rangle)(\langle BC\rangle-\langle B\rangle\langle
C\rangle)(\langle AC\rangle-\langle A\rangle\langle C\rangle)|.
\end{array}
\eqno(4.1)$$ Particularly, for the case when
$\Delta_A\Delta_C=|\langle AC\rangle-\langle A\rangle\langle
C\rangle|$ or $\langle AB\rangle=\langle A\rangle\langle B\rangle$
or $\dim H=2$,
$$
\Delta_A\Delta_B\Delta_C\geq \frac{1}{2}(\Delta_A|\langle
BC\rangle-\langle B\rangle\langle C\rangle|+\Delta_C|\langle
AB\rangle-\langle A\rangle\langle B\rangle|). \eqno(4.2)$$}

The inequalities (4.1) and (4.2) are tight as illustrated by
applying to Pauli matrices.

{\bf Example 4.2}. Uncertainty relations for Pauli matrices.

Let $X,Y,Z$ be Pauli matrices, that is,
$$X =  \begin{pmatrix}
0 & 1 \cr 1 & 0 \cr
\end{pmatrix},  \qquad
Y =\begin{pmatrix} 0 & -i \cr i & 0 \cr
\end{pmatrix},  \qquad
Z = \begin{pmatrix} 1 & 0 \cr 0 & -1 \cr
\end{pmatrix}.
$$

Recall that, for any  dense matrix $\rho\in M_2({\mathbb C})$,
$\rho$ has a representation
$$\rho=\frac{1}{2}(I_2+r_1X+r_2Y+r_3Z)$$ with Bloch vector
$(r_1,r_2,r_3)^t\in{\mathbb R}^3$  and $r_1^2+r_2^2+r_3^2\leq 1$;
and $\rho$ is pure if and only if $r_1^2+r_2^2+r_3^2= 1$. Recall
also  that $XY=iZ$,  $YZ=iX$ and $\Delta_A^2=1-\langle A\rangle^2$
for $A\in\{X,Y,Z\}$, $\langle X^2\rangle=\langle Y^2\rangle=\langle
Z^2\rangle=1$ and $(\langle X\rangle,\langle Y\rangle,\langle
Z\rangle)=( r_1, r_2, r_3)$.

Applying the inequality (4.2) of Theorem 4.1 to $X,Y,Z$ we get
$$\begin{array}{rl}
 & \Delta_X\Delta_Y\Delta_Z \\ \geq &
\frac{1}{2}(\Delta_X|i\langle X\rangle-\langle Y\rangle\langle
Z\rangle|+\Delta_Z|i\langle
Z\rangle-\langle X\rangle\langle Y\rangle|)\\
=& \frac{1}{2}(\sqrt{(1-\langle X\rangle^2)( \langle
X\rangle^2+\langle Y\rangle^2\langle Z\rangle^2)} +\sqrt{(1-\langle
Z\rangle^2)( \langle Z\rangle^2+\langle X\rangle^2\langle
Y\rangle^2)}\ ).
\end{array}
\eqno(4.3)$$

Obviously, the  inequality (4.3) is   tight and ``=" holds if the
Bloch vector satisfies $|r_1|=|r_3|=\frac{1}{\sqrt{2}}$ and $r_2=0$.

{\it This illustrates that Theorem 2.2 is tight for three
observables. Since  Schr\"{o}dinger's uncertainty principle (1.3) is
tight and our uncertainty relation is equivalent to
Schr\"{o}dinger's uncertainty principle by (3.2), Theorem 2.2 is
also tight for two obserables.}

Moreover, by Theorem 3.1,
$$(1-\langle
Z\rangle^2)(1-\langle X\rangle^2)\geq \langle Y\rangle ^2+\langle
X\rangle^2\langle Z\rangle^2,
$$
hence we have $$\begin{array}{rl}
 & \Delta_X^2\Delta_Y^2\Delta_Z^2 \\
 \geq & \frac{1}{4}[(1-\langle
X\rangle^2)( \langle X\rangle^2+\langle Y\rangle^2\langle
Z\rangle^2)+(1-\langle Z\rangle^2)( \langle
Z\rangle^2+\langle X\rangle^2\langle Y\rangle^2)]\\
& +\frac{1}{2}\sqrt{(1-\langle Z\rangle^2)(1-\langle X\rangle^2)(
\langle X\rangle^2+\langle Y\rangle^2\langle Z\rangle^2)( \langle
Z\rangle^2+\langle X\rangle^2\langle Y\rangle^2)}\\
\geq & \frac{1}{4}[(1-\langle X\rangle^2)( \langle
X\rangle^2+\langle Y\rangle^2\langle Z\rangle^2)+(1-\langle
Z\rangle^2)( \langle
Z\rangle^2+\langle X\rangle^2\langle Y\rangle^2)]\\
& +\frac{1}{2}\sqrt{( \langle Y\rangle ^2+\langle X\rangle^2\langle
Z\rangle^2)( \langle X\rangle^2+\langle Y\rangle^2\langle
Z\rangle^2)( \langle
Z\rangle^2+\langle X\rangle^2\langle Y\rangle^2)}\\
\geq & \sqrt{( \langle Y\rangle ^2+\langle X\rangle^2\langle
Z\rangle^2)( \langle X\rangle^2+\langle Y\rangle^2\langle
Z\rangle^2)( \langle Z\rangle^2+\langle X\rangle^2\langle
Y\rangle^2)}\\
\geq& 2\sqrt{2}|\langle X\rangle\langle Y\rangle\langle
Z\rangle|^{\frac{3}{2}}.\end{array} \eqno(4.4)$$

Particularly, one has
$$\Delta_X^2\Delta_Y^2\Delta_Z^2\geq  2\sqrt{2}|\langle X\rangle\langle Y\rangle\langle
Z\rangle|^{\frac{3}{2}}. \eqno(4.5)$$

Observe that we always have
$$1\geq\Delta_X^2\Delta_Y^2\Delta_Z^2=
 (1- r_1^2)(1- r_2^2)(1- r_3^2) \geq
\frac{8}{27}$$ since the function $(1- r_1^2)(1- r_2^2)(1- r_3^2)$
has its minimum value $\frac{8}{27}$ at
$|r_1|=|r_2|=|r_3|=\frac{1}{\sqrt{3}}$ and the maximum value 1 at
$\rho=\frac{1}{2}I_2$. Moreover,
  $|r_1r_2r_3|$ achieves simultaneously its maximum value
$\frac{1}{3\sqrt{3}}$ at $|r_1|=|r_2|=|r_3|=\frac{1}{\sqrt{3}}$.
Thus the inequality (4.5) can be sharped to
$$
\Delta_X^2\Delta_Y^2\Delta_Z^2 \geq \frac{8\sqrt[4]{3}}{3}|\langle
X\rangle\langle Y\rangle\langle Z\rangle|^{\frac{3}{2}}.
\eqno(4.6)$$

The inequality (4.6) is tight in the sense that ``=" holds if
$|r_1|=|r_2|=r_3|=\frac{1}{\sqrt{3}}$.

Compare  (4.3) with (1.6) and (4.6). Although these inequalities are
all tight, one of the remarkable advantage of (4.3) is that, even if
some of $\langle X\rangle,\langle Y\rangle,\langle Z\rangle$ are
zero, we still may get a positive lower bound of $\Delta_X \Delta_Y
\Delta_Z$. For instance, saying $\langle Y\rangle=0$, we have
$$\Delta_X \Delta_Y \Delta_Z \geq
\frac{1}{2 }\left(\sqrt{(1-\langle X\rangle^2) \langle X\rangle^2}
+\sqrt{(1-\langle Z\rangle^2)\langle Z\rangle^2}\right);
$$
saying  $\langle Y\rangle=\langle Z\rangle=0$, we have
$$\Delta_X \Delta_Y \Delta_Z \geq
\frac{1}{2 } \sqrt{(1-\langle X\rangle^2) \langle X\rangle^2},$$
while we cannot get any information from (1.6) and (4.6).

Before conclusion we state the uncertainty relation from Theorem 2.2
for four observations, which has a relatively simple expression.

{\bf Theorem 4.3}. {\it Let $A_1,A_2,A_3,A_4$ be observables. Then
$$\begin{array}{rl}
& \Delta_{A_1}\Delta_{A_2}\Delta_{A_3}\Delta_{A_4}\\
\geq & \frac{1}{2} |\langle A_2A_3\rangle-\langle A_2\rangle\langle
A_3\rangle|(|\langle A_1A_4\rangle-\langle A_1\rangle\langle
A_4\rangle|+ \Delta_{A_1}\Delta_{A_4}).
\end{array} \eqno(4.7)$$}

The inequality (4.7) is tight. For example, Consider bipartite
continuous-variable system. Let $(A_1, A_4,
A_2,A_3)=(\hat{q}_1,\hat{p}_1,\hat{q}_2,\hat{p}_2)$, where
$\hat{q}_i,\hat{p}_i$ are the position and momentum in the $i$th
mode satisfying the canonical commutation relation. As Heisenberg's
uncertainty principle (1.1) is tight, we say that
$$\begin{array}{rl}
& \Delta_{{q}_1}\Delta_{{p}_1}\Delta_{{q}_2}\Delta_{{p}_2}\\
\geq & \frac{1}{2} |\langle \hat{q}_2\hat{p}_2\rangle-\langle
\hat{q}_2\rangle\langle \hat{p}_2\rangle|(|\langle
\hat{q}_1\hat{p}_1\rangle-\langle \hat{q}_1\rangle\langle
\hat{p}_1\rangle|+ \Delta_{q_1}\Delta_{p_4}).
\end{array} \eqno(4.8)$$
is tight, the ``$=$" is attained at $\rho=e$.

Similarly, considering the   positions and momentums
$(\hat{q}_1,\hat{p}_1,\hat{q}_2,\hat{p}_2,\ldots,
\hat{q}_n,\hat{p}_n)$ in a $n$-partite continuous-variable system,
one sees that the uncertainty relation (2.4) in Theorem 2.2 is
tight.

However we do not know whether   the uncertainty relation (2.5) is
tight for odd $k=2n+1\geq 5$.

\section{Conclusion}

Uncertainty relations discover lower bounds of the product of
standard deviations of several observables. Larger the lower bound
is, more powerful the corresponding uncertainty relation is. There
are no known uncertainty relations that valid for arbitrary multi
observavles.
 By
finding the equality of deviation and the norm of the Lie product of
the observable and the pure state,  we reduce the question of
establishing uncertainty relation of multi observerbles to the
question of computing the numerical radius of an operator of rank
$\leq 2$. This enable us establish a universal  uncertainty relation
for any $k$  observables, of which, the formulation depends on the
even or odd quality of $k$. For two observables, our uncertainty
relation is exactly a simpler reformulation of Schr\"{o}dinger's
uncertainty principle. The uncertainty relation provided in this
paper is tight, at least for the cases of two and three observables,
as illustrated by   examples.

\section{Appendix}

In the appendix, we give the proofs of theorems 2.1 and 2.2.

\begin{proof}[Proof of Theorem 2.1] Let $H$ be the associated Hilbert space for the  pure state $|x\rangle$ and the observable
$A$. Write $A|x\rangle$ in the form $A|x\rangle=\alpha
|x\rangle+\beta |y\rangle$, where normalized $|y\rangle$ is
orthogonal to $|x\rangle$. Since $A$ is self-adjoint we have
$\alpha=\langle x|A|x \rangle\in\mathbb R$. Moreover, by
self-adjointness of $A$, the Lie product of $A$ and the rank one
projection $ |x\rangle\langle x|$ is represented by the following
matrix relative to decomposition $H=[x]\oplus [y]\oplus
\{x,y,\}^\perp$, here $[x]={\rm span}\{x\}$. Then
$$[A, |x\rangle\langle x|]=\left(\begin{matrix}
0 & -\bar{\beta} \\
\beta& 0
\end{matrix}\right)\oplus 0.$$
Note that $[A, |x\rangle\langle x|]$ is a  skew self-adjoint
operator because $[A, |x\rangle\langle x|]^\dag=-[A,
|x\rangle\langle x|]$. Thus its numerical range   $W([A,
|x\rangle\langle x|])=i[-|\beta|,|\beta|]$,  and hence $w([A,
|x\rangle\langle x|])=\|[A,x\otimes x]\|=|\beta|$. It follows that
$$\begin{array}{lllll}w([A, |x\rangle\langle x|])^2& =\|[A, |x\rangle\langle x|]\|=|\beta|^2 \\
&=\|A|x\rangle-\langle x|A|x\rangle\, |x\rangle\|^2 \\ & = \langle
x|(A -\langle x|A|x\rangle)^2 |x\rangle
\\ & =\langle x|A^2|x\rangle-(\langle x|A|x\rangle )^2 \\ &= \langle A^2 \rangle-\langle A\rangle^2= \Delta_A^2. \end{array}$$
Therefore,  $\Delta_A=w([A, |x\rangle\langle x|])=\|[A,
|x\rangle\langle x|]\|$, completing the proof. \end{proof}

Before start the proof of Theorem 2.2, we need a lemma.

{\bf Lemma A1}. {\it Let
$$E_1=\left(\begin{array}{ccc} 0 &a & b \\ c & 0 &0\\ 0 & 0 & 0
\end{array}\right), E_2=\left(\begin{array}{cc} 0 &a \\
c&0\end{array}\right).$$ Then
$$
w(E_1)=\frac{1}{2}\sqrt{|b|^2+(|a|+|c|)^2}
$$ and
$$
w(E_2)=\frac{1}{2}(|a|+|c|).$$ }

\begin{proof}
Obviously, with $ac=|ac|e^{2i\theta}$, $\sigma (E_1)=\{\pm
\sqrt{|ac|}e^{i\theta},0\}$. It is easily checked that $E_1$
satisfies the conditions of Theorem 2.3 and 2.4 of Ref \cite{KRS},
and hence the numerical range $W(E_1)$ of $E_1$ is an elliptic
disc with foci $\{\pm \sqrt{|ac|}e^{i\theta}\}$. Thus the
numerical radius $w(E_1)$ is the half length of major axis of the
ellipse.

Let $F=e^{-i\theta} E_1$; then $w(F)=w(E_1)$. As $\sigma(F)=\{\pm
\sqrt{|ac|},0\}$, we see that
$$w(F)=\|{\rm Re}(F)\|.$$ Note that
$$\sigma({\rm Re}(F))=\{ 0,
\pm\frac{1}{2}\sqrt{|b|^2+|ae^{-i\theta}+\bar{c}e^{i\theta}|^2}\
\}.$$ A simple computation shows that
$$|ae^{-i\theta}+\bar{c}e^{i\theta}|^2=(|a|+|c|)^2.$$
Therefore,  we have
$$
w(E_1)=w(F)= \frac{1}{2}\sqrt{|b|^2+(|a|+|c|)^2}.
$$
 It is also easily checked that
 $$w(E_2)= \frac{1}{2}|ae^{-i\theta}+\bar{c}e^{i\theta}|= \frac{1}{2}(|a|+|c|).$$
\end{proof}

\begin{proof}[Proof of Theorem 2.2] As remarked after the statement
os Theorem 2.2, it is enough to prove the theorem for the pure
states.

For any given observables $A_1,A_2, \ldots, A_k$ with $k\geq 2$, let
$D_k=\Pi_{j=1}^k[A_j,|x\rangle\langle x|]$, which has the form
$$D_k=a_k|x\rangle\langle x|+b_k|A_1x\rangle\langle x|+c_k
|x\rangle\langle xA_k|+d_k|A_1x\rangle\langle xA_k|.$$ A direct
computation gives $$ \left\{ \begin{array}{l}
a_2=-\langle A_1A_2\rangle,\\ b_2= \langle A_2\rangle, \\
c_2=\langle A_1\rangle,\\ d_2=-1.\end{array}\right.\eqno(A1)$$
 For any $k\geq 3$, since $D_k=D_{k-1}[A_k, |x\rangle\langle x|]$, one may
 take
$$\left\{ \begin{array}{l}
a_k=a_{k-1}\langle A_k\rangle+c_{k-1}\langle A_{k-1}A_k\rangle,\\
b_k=b_{k-1}\langle A_k\rangle+d_{k-1}\langle A_{k-1}A_k\rangle,\\
c_k=-a_{k-1}-c_{k-1}\langle A_{k-1}\rangle,\\
d_k=-b_{k-1}-d_{k-1}\langle
A_{k-1}\rangle.\end{array}\right.\eqno(A2)$$

Take unitors vectors $|y\rangle,|z\rangle$ so that
$\{|x\rangle,|y\rangle,|z\rangle\}$ is orthogonal and
$$ \left\{\begin{array}{l}
|A_1x\rangle=\langle A_1\rangle|x\rangle+\Delta_{A_1}|y\rangle,\\
|A_kx\rangle=\langle A_k\rangle|x\rangle +\beta^\prime
|y\rangle+\gamma^\prime |z\rangle.
\end{array}\right.$$
Then
$$\beta^\prime= \Delta_{A_1}^{-1}(\langle A_1A_k\rangle-\langle
A_1\rangle\langle A_k\rangle),\quad
 \Delta_{A_k}=\sqrt{|\beta^\prime|^2+|\gamma^\prime|^2}. \eqno(A3)$$ and
$$
\begin{array}{rl}
D_k=& (a_k+b_k\langle A_1\rangle+c_k\langle A_k\rangle+d_k\langle
A_1\rangle\langle A_k\rangle)|x\rangle\langle x|+(c_k+d_k\langle
A_1\rangle)\bar{\beta^\prime}|x\rangle\langle
y|\\
&+(c_k+d_k\langle A_1\rangle)\bar{\gamma^\prime}|x\rangle\langle
z|+(b_k+d_k\langle A_k\rangle)\Delta_{A_1}|y\rangle\langle x|\\
&+d_k\Delta_{A_1}\bar{\beta^\prime}|y\rangle\langle
y|+d_k\Delta_{A_1}\bar{\gamma^\prime}|y\rangle\langle z|\\
=& f_{11}^{(k)}|x\rangle\langle x|+f_{12}^{(k)}|x\rangle\langle
y|+f_{13}^{(k)}|x\rangle\langle z| \\ &
+f_{21}^{(k)}|y\rangle\langle x|+f_{22}^{(k)}|y\rangle\langle
y|+f_{23}^{(k)}|y\rangle\langle z|.
\end{array}\eqno(A4)$$
Note that, by (A2), we have
$$ \begin{array}{rl} f_{11}^{(k)}=&
c_{k-1}(\langle A_{k-1}A_{k}\rangle-\langle A_{k-1}\rangle\langle
A_k\rangle)+d_{k-1}(\langle A_1\rangle\langle
A_{k-1}A_k\rangle-\langle A_1\rangle\langle A_{k-1}\rangle\langle
A_k\rangle)\\
=& -f_{11}^{(k-2)}(\langle A_{k-1}A_{k}\rangle-\langle
A_{k-1}\rangle\langle A_k\rangle) \end{array}
$$
as
$$\begin{array}{rl} & c_k+d_k\langle A_1\rangle \\
=&- c_{k-2}(\langle A_{k-2}A_{k-1}\rangle-\langle
A_{k-2}\rangle\langle A_{k-1}\rangle)-d_{k-2}(\langle
A_1\rangle\langle A_{k-2}A_{k-1}\rangle-\langle A_1\rangle\langle
A_{k-2}\rangle\langle A_{k-1}\rangle)\\=
&-f_{11}^{(k-1)},\end{array}
$$
$$d_k=-b_{k-1}-d_{k-1}\langle A_{k-1}\rangle =d_{k-2}(\langle A_{k-2}\rangle\langle A_{k-1}\rangle-\langle A_{k-2}A_{k-1}\rangle) $$
and
$$ b_k+d_k\langle A_k\rangle=d_{k-1}(\langle
A_{k-1}A_k\rangle-\langle A_{k-1}\rangle\langle
A_k\rangle)=-d_{k+1},
$$
 which reveal that
$$\begin{array}{l} f_{11}^{(k)}=-f_{11}^{(k-2)}(\langle
A_{k-1}A_{k}\rangle-\langle
A_{k-1}\rangle\langle A_k\rangle),\\
f_{12}^{(k)}=-f_{11}^{(k-1)}\bar{\beta^\prime},\\
f_{13}^{(k)}=-f_{11}^{(k-1)}\bar{\gamma^\prime},\\
 f_{21}^{(k)}=-d_{k+1}\Delta_{A_1}=(\langle
A_{k-1}A_k\rangle-\langle A_{k-1}\rangle\langle
A_k\rangle)f_{21}^{(k-2)},  \\
f_{22}^{(k)}=d_k\Delta_{A_1}\bar{\beta^\prime}=(\langle
A_{k-2}A_{k-1}\rangle-\langle A_{k-2}\rangle\langle
A_{k-1}\rangle)d_{k-2}\Delta_{A_1}\bar{\beta^\prime}, \\
f_{23}^{(k)}=d_k\Delta_{A_1}\bar{\gamma^\prime}=(\langle
A_{k-2}A_{k-1}\rangle-\langle A_{k-2}\rangle\langle
A_{k-1}\rangle)d_{k-2}\Delta_{A_1}\bar{\gamma^\prime}.
\end{array}
\eqno(A5)$$

It is easily checked that
$$D_2=\begin{array}{ccc}&\left(\begin{matrix}
\langle A_1\rangle\langle A_2\rangle-\langle  A_1A_2  \rangle & 0 &0 \\
0 & -\Delta_{A_1} \beta' & -\Delta_{A_1} \gamma' \\
0 & 0 & 0\\
\end{matrix}\right)\oplus 0\end{array}$$
and
$$
D_3=\left(\begin{array}{ccc} 0 &(\langle A_1A_2\rangle-\langle
A_1\rangle\langle A_2\rangle)\beta^\prime & (\langle
A_1A_2\rangle-\langle A_1\rangle\langle A_2\rangle)\gamma^\prime \\
(\langle A_2\rangle\langle A_3\rangle -\langle A_2A_3\rangle)\Delta_{A_1} & 0 &0\\
0 & 0 & 0
\end{array}\right) \oplus 0
$$
if $\dim H\geq 3$;
$$
D_2=\left(\begin{array}{cc} \langle A_1\rangle\langle
A_2\rangle-\langle  A_1A_2  \rangle &0
 \\ 0 & -\Delta_{A_1} \Delta_{A_2}
\end{array}\right)$$ and $$ D_3=\left(\begin{array}{cc} 0 &(\langle A_1A_2\rangle-\langle
A_1\rangle\langle A_2\rangle)\Delta_{A_2}
 \\ (\langle A_2\rangle\langle A_3\rangle -\langle A_2A_3\rangle)\Delta_{A_1} & 0
\end{array}\right)
$$
if $\dim H=2$. Hence, by identify the case of $\dim H=2$ with the
case $\gamma^\prime=0$, we may agree that $D_2$ and $D_3$ have
respectively the patten
$$
\left(\begin{array}{ccc} * & 0&0\\ 0 & * & * \\ 0&0&0
\end{array}\right)\oplus 0
\eqno(A6)$$
 and
$$
\left(\begin{array}{ccc} 0 &*&*\\ *& 0& 0 \\ 0&0&0
\end{array}\right)\oplus 0.
\eqno(A7)$$
 Then Eq.(A5) implies that  $D_{2n}$ has the patten
 (A6) if $k=2n$ is even and $D_{2n+1}$ has the patten  (A7)) if $k=2n+1$ is odd.

Let us first calculate $w(D_2)$. It is easily checked that
$$\Delta_{A_1}\Delta_{A_2}\geq w(D_2)= \max\{|\langle A_1A_2 \rangle-\langle A_1\rangle\langle
A_2 \rangle|,\frac{1}{2}|\langle A_1A_2 \rangle-\langle A_1 \rangle
\langle A_2 \rangle|+\frac{\Delta_{A_1}\Delta_{A_2}}{2}\}.$$ Thus we
have
$$\Delta_{A_1}\Delta_{A_2}\geq |\langle A \rangle\langle B \rangle-\langle AB
\rangle| \eqno(A8)$$ and
$$w(D_2)=  \frac{1}{2}(|\langle A_1A_2 \rangle-\langle A_1 \rangle
\langle A_2 \rangle|+ {\Delta_{A_1}\Delta_{A_2}}) .$$

If $k=2n$ is even, then it follows from (A5) that
$$|f_{11}^{(2n)}|=|(\langle A_{1}A_{2}\rangle-\langle
A_{1}\rangle\langle A_1\rangle)\cdots (\langle
A_{k-1}A_{k}\rangle-\langle A_{k-1}\rangle\langle A_k\rangle)| .
\eqno(A9)
$$
Since $D_{2n}$ has patten (A6), by (A5) and (A8), we have
$$\|D_{2n}\|=\max\{|f_{11}^{(2n)}|,\sqrt{|f_{22}^{(2n)}|^2+|f_{23}^{(2n)}|^2}\ \}=|d_{2n}|\Delta_{A_1}\Delta_{A_{2n}}$$
and
$$ \begin{array}{rl}
w(D_{2n})=&\max\{|f_{11}^{(2n)}|,\frac{1}{2}\left(|f_{22}^{(2n)}|+\sqrt{|f_{22}^{(2n)}|^2+|f_{23}^{(2n)}|^2}\right)\}\\
=& \frac{1}{2}|d_{2n}|(|\langle A_1A_{2n}\rangle-\langle
A_1\rangle\langle A_{2n}\rangle|+\Delta_{A_1}\Delta_{A_{2n}}).
\end{array}$$
Now $d_2=-1$ and $$|d_{2n}|=|d_{2n-2}|\cdot|\langle
A_{2n-2}A_{2n-1}\rangle-\langle A_{2n-2} \rangle\langle
A_{2n-1}\rangle|$$ entails that
$$ \begin{array}{rl} w(D_{2n})=
\frac{1}{2}(\Pi_{j=1}^{n-1}|\langle A_{2j}A_{2j+1}\rangle-\langle
A_{2j } \rangle\langle
 A_{2j+1}\rangle|)(|\langle A_1A_{2n}\rangle-\langle A_1\rangle\langle
A_{2n}\rangle|+\Delta_{A_1}\Delta_{A_{2n}}).
\end{array}\eqno(A10)$$
As $\Pi_{j=1}^{2n}\Delta_{A_j}\geq w(D_{2n})$, this completes the
proof of (2.4).

If $k=2n+1$ is odd, then $D_{2n+1}$ has the patten (A7). Applying
Lemma A1 gives
$$\begin{array}{rl}
w(D_{2n+1})=&\frac{1}{2}\sqrt{(|f_{12}^{(2n+1)}|+|f_{21}^{(2n+1)}|)^2+|f_{13}^{(2n+1)}|^2}\\
=&\frac{1}{2}\sqrt{(|f_{11}^{(2n)}\beta^\prime|+|d_{(2n+2)}|\Delta_{A_1})^2+|f_{11}^{(2n)}\gamma^\prime|^2}\\
=&\frac{1}{2}\sqrt{(2|f_{11}^{(2n)}d_{(2n+2)}|\Delta_{A_1}\beta^\prime|+|d_{(2n+2)}|^2\Delta_{A_1}^2+|f_{11}^{(2n)}|^2(|\beta^\prime|^2+|\gamma^\prime|^2)}.
\end{array}$$
Therefore, $$ w(D_{2n+1}) =\frac{1}{2}\sqrt{2\pi_1\pi_2|\langle
A_1A_{2n+1}\rangle-\langle A_1\rangle\langle
A_{2n+1}\rangle|+\pi_2^2\Delta_{A_1}^2+\pi_1^2\Delta_{A_{2n+1}}^2},
\eqno(A11)$$
 where
$$\pi_1=\Pi_{j=1}^{n}|\langle A_{2j-1}A_{2j}\rangle-\langle A_{2j-1}\rangle\langle
A_{2j}\rangle|$$ and
$$\pi_2=\Pi_{j=1}^{n}|\langle A_{2j}A_{2j+1}\rangle-\langle A_{2j}\rangle\langle
A_{2j+1}\rangle|.$$ As $\Pi_{j=1}^{2n+1}\Delta_{A_j}\geq
w(D_{2n+1})$,  we complete the proof of (2.5) by  (A11).
\end{proof}

Finally we explain why   the sharper inequality
$$\Pi_{j=1}^k\Delta_{A_j}\geq\|D_k\| \eqno(A12)$$ cannot
achieve sharper uncertainty relations than the weaker inequality
$\Pi_{j=1}^k\Delta_{A_j}\geq w(D_k)$ can.

If $k=2n$ is even, then by (A5) and (A6) one has
$$\|D_{2n}\|=\Delta_{A_1}\Delta_{A_{2n}}\Pi_{j=1}^{n-1}|\langle A_{2j}A_{2j+1}\rangle-\langle
A_{2j } \rangle\langle
 A_{2j+1}\rangle|.$$
 This together with (A12) gives
$$\Pi_{j=2}^{2n-1}\Delta_{A_j}\geq \Pi_{j=1}^{n-1}|\langle A_{2j}A_{2j+1}\rangle-\langle
A_{2j } \rangle\langle
 A_{2j+1}\rangle|,$$
 which is weaker than the inequality (2.4).

 If $k=2n+1$ is odd, then by (A5) and (A7) one has
 $$\begin{array}{rl} \|D_{2n+1}\|=& \max\{ \sqrt{|f_{12}^{(2n+1)}|^2+|f_{13}^{(2n+1)}|^2},
 |f_{21}^{(2n+1)}|\}\\
 =&\max\{ |f_{11}^{(2n)}|\Delta_{A_{2n+1}},
 |d_{2n+2}|\Delta_{A_1}\},
 \end{array}$$
which gives
$$\Pi_{j=1}^{2n}\Delta_{A_j}\geq \Pi_{j=1}^{n}|\langle A_{2j-1}A_{2j}\rangle-\langle
A_{2j-1 } \rangle\langle
 A_{2j}\rangle|$$
 or
$$\Pi_{j=2}^{2n+1}\Delta_{A_j}\geq  \Pi_{j=1}^{n}|\langle
A_{2j}A_{2j+1}\rangle-\langle A_{2j } \rangle\langle
 A_{2j+1}\rangle|,$$
again weaker than (2.4).

{\bf Acknowledgements} This work is partly supported by National
Science Foundation of China (11201329, 11171249) and Program for the
Outstanding Innovative Teams of Higher Learning Institutions of
Shanxi.

{\bf Competing interests statement} The authors declare that they
have no competing financial interests.

{\bf Correspondence}  should be addressed to J. C. Hou
(houjinchuan@tyut.edu.cn).

\end{document}